\documentstyle[aps,epsf,floats]{revtex}

\begin{document}

\baselineskip 14pt

\title{The Detection of Gravitational Waves with LIGO}
\author{Barry C. Barish}
\address{California Institute of Technology, Pasadena, CA 91125}
\maketitle

\begin{abstract}
Gravitational wave emission is expected to arise from a variety of
astrophysical phenomena.  A new generation of detectors with
sensitivity consistent with expectation from such sources is being
developed.  The Laser Interferometer Gravitational-Wave Observatory
(LIGO), one of these ambitious undertakings, is being developed by a
Caltech-MIT collaboration.  It consists of two widely separated
interferometers, which will be used in coincidence to search for
sources from compact binary systems, spinning neutron stars,
supernovae and other astrophysical or cosmological phenomena that emit
gravitational waves.  The construction of LIGO is well underway and
preparations are being made for the commissioning phase.  In this
lecture, I review the underlying physics of gravitational waves,
review possible astrophysical and cosmological sources and discuss the
LIGO interferometer status and plans.
\end{abstract}

\section {Introduction}

Einstein first predicted gravitational waves in 1916 as a consequence
of the general theory of relativity.  In this theory, concentrations
of mass (or energy) warp space-time, and changes in the shape or position 
of such objects cause a distortion that propagates through the Universe at the
speed of light ({\it i.e.}, a gravitational wave).

It is tempting to draw the analogy between gravitational waves and
electromagnetic waves.  However, the nature of the waves is quite
different in these two cases.  Electromagnetic waves are oscillating
electromagnetic fields propagating through space-time, while
gravitational waves are the propagation of distortions of space-time,
itself.  The emission mechanisms are also quite different.
Electromagnetic wave emission results from an incoherent superposition
of waves from molecules, atoms and particles, while gravitational
waves are coherent emission from bulk motions of energy.  The
characteristics of the waves are also quite different in that
electromagnetic waves experience strong absorption and scattering in
interaction with matter, while gravitational waves have essentially no
absorption or scattering.  Finally, the typical frequency of detection
of electromagnetic waves is $f > 10^7$ Hz, while gravitational waves are
expected to be detectable at much lower frequency, $f < 10^4$ Hz.

By making these comparisons it becomes clear that most sources of
gravitational waves will not be seen as sources of electromagnetic
waves and vice versa.  This means there is great potential for
surprises! However, it also means that (since most of what we know
about the Universe comes from electromagnetic waves) there is much
uncertainty in the types and characteristics of sources, as well as
the strengths and rate.

\begin{figure}[ht]
\centerline{\epsfysize=8cm\epsfbox{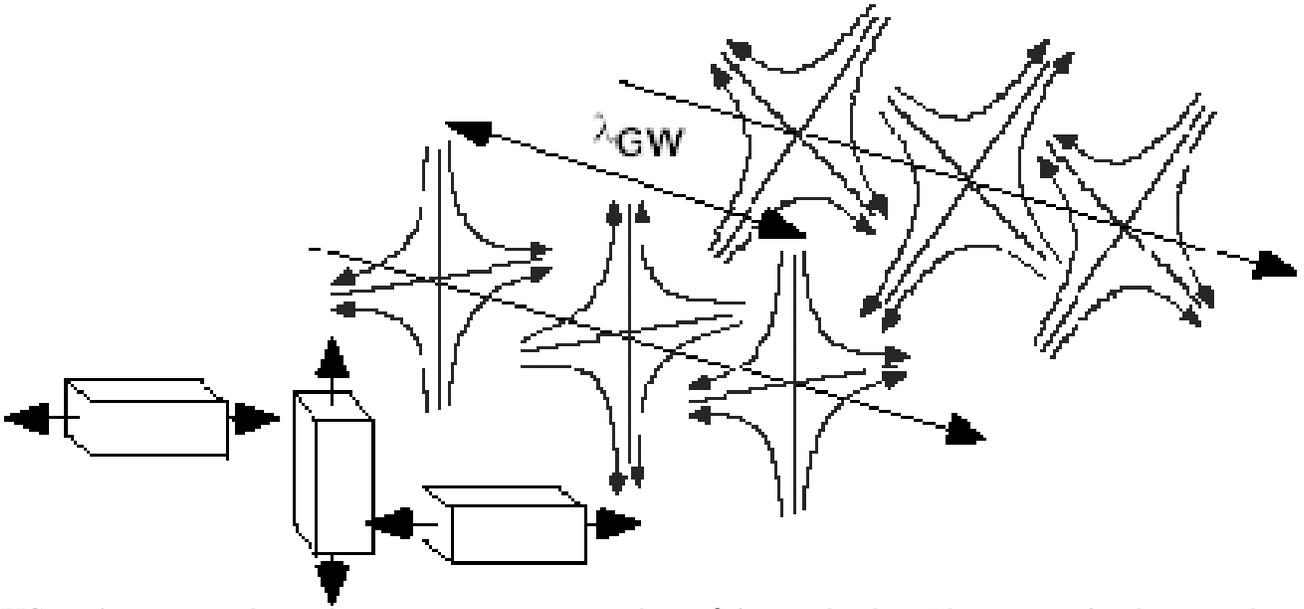}}
%\vskip -.2 cm
\caption[]{\label{fig1}\small
A gravitational wave has two components oriented at $45^\circ$ from
each other.  The passage of such a wave distorts space-time and
produces change in length in two orthogonal directions which oscillate
with gravitational wave frequency.}
\end{figure}

The characteristics of gravitational radiation can be seen from the
perturbation to flat space-time and in the weak field approximation is
expressed by
$ g_{\mu \nu}= \eta_{\mu \nu} + h_{\mu \nu} ,$
where $h_{\mu \nu}$ is the perturbation from Minkowski space.  The detailed
information about the gravitational wave is carried in the form of the
quantity $h_{\mu \nu}$.
There is freedom of the choice of gauge, but in the
transverse traceless gauge and weak field limit, the field equations
become a wave equation
$$ \left( \nabla^2 - \frac{1}{c^2} 
          \frac{\partial^2\;}{\partial t^2} \right) h_{\mu \nu} = 0, $$
with the solution being plane waves having two polarizations for the
gravitational wave,
$$ h_{\mu \nu} = a \hat{h}_+\left(t - \frac{z}{c}\right)
               + b \hat{h}_\times\left(t - \frac{z}{c}\right),$$
with the two components at $45^\circ$ from each other (Figure \ref{fig1}),
rather than $90^\circ$ as for electromagnetic waves.

Interestingly, this is a consequence of the spin 2 nature of gravity.
The experiments discussed below, although classical experiments
analogous to the Hertz experiment that demonstrated electromagnetic
waves, are capable of decomposing the two components of the wave,
thereby establishing empirically that gravity is spin 2.  They also
have capability to measure the speed of the gravitational wave, and can
establish that they move with velocity $c$.

For gravitational radiation there is no monopole term or dipole term,
so the first term is quadrupolar and the
strength of the radiation depends on the magnitude of this non
axisymmetric moment.  The largest term for gravitational radiation is
$$ h_{\mu\nu} = \frac{2G}{Rc^4} \ddot{I}_{\mu\nu},$$
where $G$ is Newton's constant, $R$ is the distance to the source, 
and $I_{\mu\nu}$ is the reduced quadrupole moment tensor.
This yields a strain at the surface of the earth for the inspiral
of a binary system of two neutron stars at a distance of the Virgo
Cluster ($\sim$15 Mpc) of $h \simeq 10^{-21}$.  
The new generation of gravitational
wave detectors promise to have resolution capable of measuring such
small strain.

\begin{figure}[ht]
\centerline{\epsfysize=8cm\epsfbox{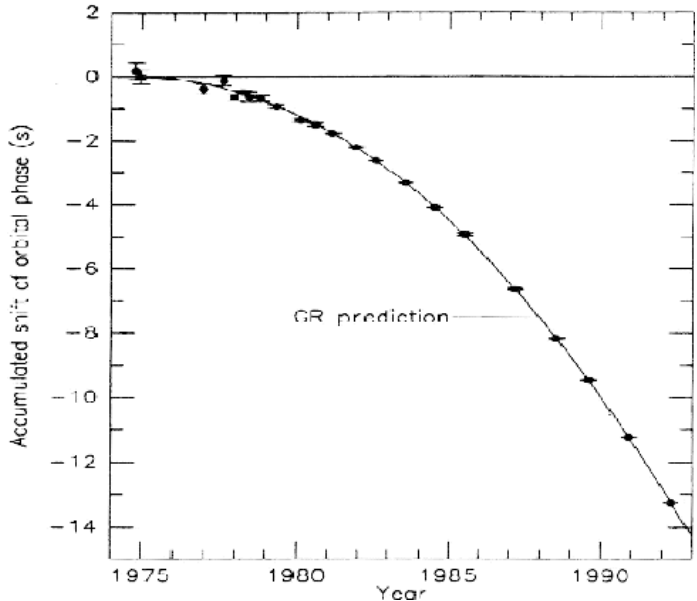}}
\smallskip
\caption[]{\label{fig2}\small
         Speed up of orbital period for PSR 1913+16 as measured by
Hulse and Taylor and predictions from general relativity using
measured parameters of this binary system.}
\end{figure}

Until now, gravitational waves have not been observed directly,
however, strong indirect evidence resulted from the beautiful
experiment of Hulse and Taylor\cite{ref1}.
They studied the neutron star binary
system PSR1913+16 and observed by using pulsar timing the gradual
speed up of the $\sim$ 8 hour orbital period of this system.
This speed up
of about 10 seconds was tracked accurately over about 14 years and the
result (Figure \ref{fig2})
is in very good quantitative agreement with the predictions of
general relativity.

Of course, the motivation for direct detection of gravitational waves
is based on the empirical desire to ```see these waves''.  However, such
studies also have enormous potential both to study the nature of
gravity in a new regime and to probe the Universe in a fundamentally
new way.  It is tempting to draw an analogy with the neutrino, where
it was ``indirectly observed'' by Pauli and Fermi in the 1930's as the
explanation for the apparent non-conservation of energy and angular
momentum in nuclear beta decay.  Decades of rich physics have
followed, which were first focussed on the goal of direct detection.
Since the direct detection by Reines and Cowan, neutrino physics has
been a rich subject, both for studies of the properties of neutrinos
themselves ({\it e.g.}, the question of neutrino mass remains an important
topic) and as an important tool to probe the constituent nature of
nucleons.

\begin{figure}[ht]
\centerline{\epsfysize=10cm\epsfbox{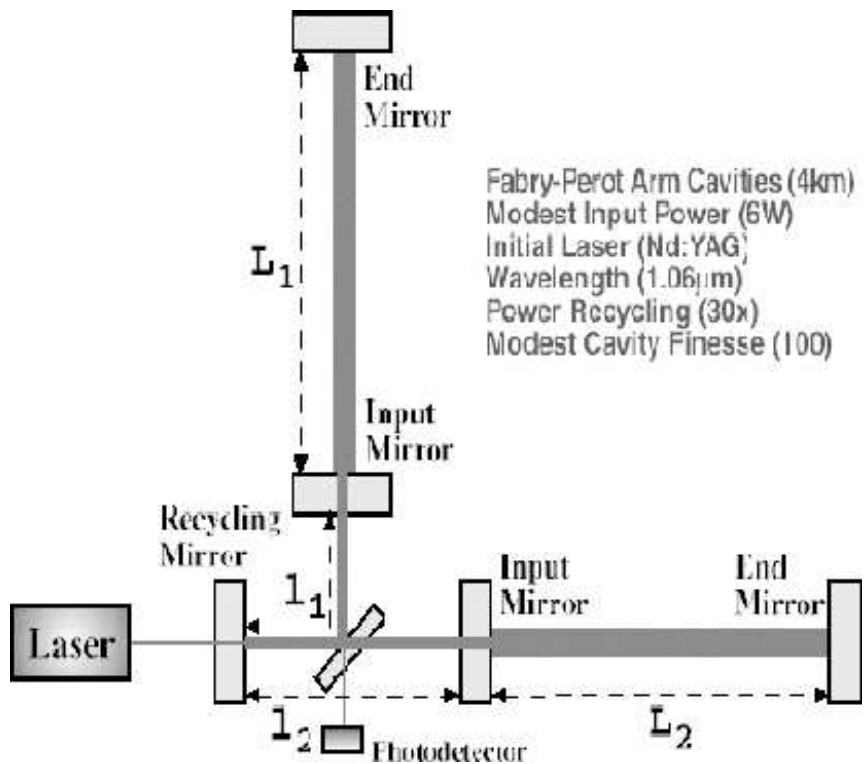}}
\smallskip
\caption[]{\label{fig3}\small
           The initial LIGO interferometer configuration.}
\end{figure}

LIGO\cite{ref2}
is designed to directly detect gravitational ways using the
technique of laser interferometry.  The arms of the interferometer are
arranged in an L-shaped pattern that will measure changes in distance
between suspended test masses at the ends of each arm.  The basic
principle is illustrated in Figure \ref{fig3}.  
A gravitational wave produces a
distortion of the local metric such that one axis of the
interferometer is stretched while the orthogonal direction shrinks.
This effect oscillates between the two arms with the frequency of the
gravitational wave.  Thus, 
$$\Delta L = \Delta L_{1} - \Delta L_{2} = hL, $$
where $h$ is the gravitational strain or amplitude of the
the gravitational wave.  Since the effect is linearly
proportional to $L$, the interferometer should have arm length as long
as is practical and for LIGO that is 4 km, to yield the target strain
sensitivity of $h \sim 10^{-21}$ for the initial interferometers now being
installed.

\section{Sources of Gravitational Waves}

Construction of LIGO is well underway at the two observatory sites:
Hanford, Washington, and Livingston, Louisiana.
The commissioning of
the detectors will begin in 2000.  The first data run is expected to
begin in 2002 at a sensitivity of $h \sim 10^{-21}$.
Incremental technical
improvements that will lead to a better sensitivity of $h \sim 10^{-22}$
are
expected to follow shortly, and the facility will allow further
improved second generation interferometers when they are developed
with sensitivity of $h \sim 10^{-23}$.
It is also important to note that all
the detectors in the world of comparable sensitivity will be used
in a worldwide network to make the most sensitive and reliable
detection.  A comparably long baseline detector (VIRGO) is being built
by a French-Italian collaboration near Pisa, and there are smaller
interferometers being built in Japan (TAMA) and in Germany (Geo-600).
Finally, an Australian group is working toward a detector in the
Southern Hemisphere.

\begin{figure}[ht]
\centerline{\epsfysize=8cm\epsfbox{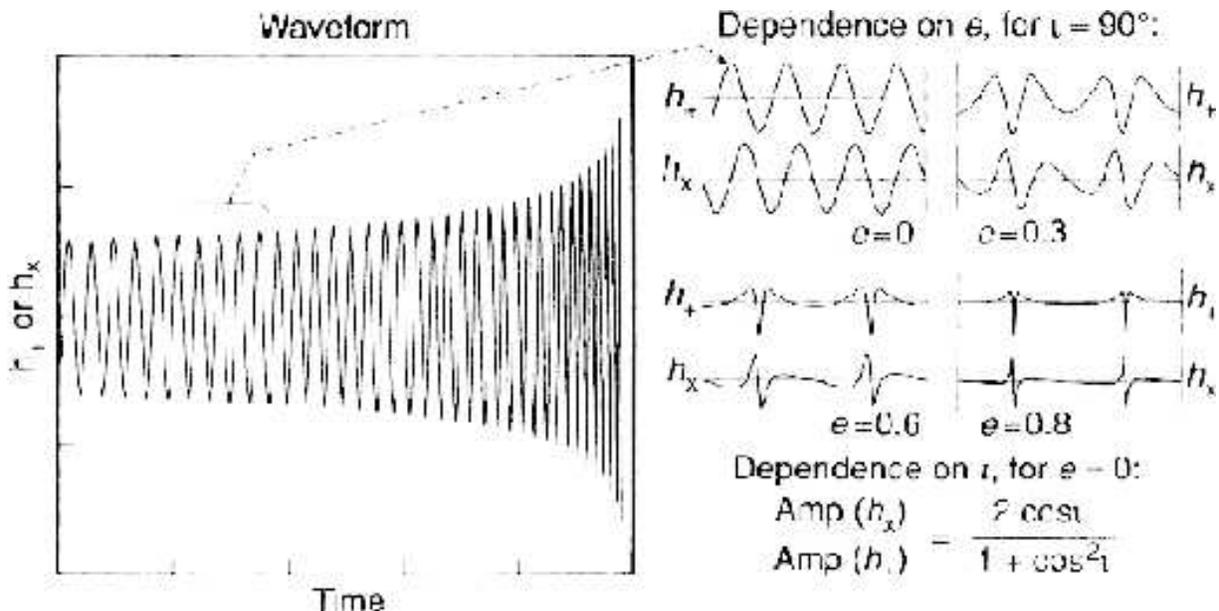}}
\smallskip
\caption[]{\label{fig4}\small
           The chirp waveform from a compact binary inspiral with
increasing amplitude and frequency is shown.  On the right are
indicated detailed waveform dependence on parameters of the binary
system: the ellipticity {\sl e}. and orbital inclination {\sl i}.}
\end{figure}

There are a large number of processes in the Universe that could emit
detectable gravitational waves. Interferometers like LIGO will
search for gravitational waves in the frequency range $f \sim 10$ Hz
to 10 KHz.
It is worth noting that there are proposals to put interferometers in
space which would be complementary to the terrestrial experiments, as
they are sensitive to much lower frequencies ( $f <$ 0.1 Hz), where there
are known sources like neutron binaries or rotating black holes.  For
LIGO, characteristic signals from astrophysical sources will be sought
by recording time-frequency data.
Examples of such signals include the following:

{\bf Chirp Signals:} The inspiral of compact objects such as a pair of
neutron stars or black holes will give radiation that increases in
amplitude and frequency as they move toward the final coalescence of
the system. This characteristic chirp signal can be characterized in
detail, depending on the masses, separation, ellipticity of the
orbits, {\it etc.}.  
Figure \ref{fig4} illustrates the ``chirp'' signal where the
amplitude and frequency are determined by the masses of the neutron
stars (the chirp mass), the distance to the sources and the orbital
inclination.  A variety of search techniques, including comparisons
with an array of templates will be used for this type of search.  The
Newtonian (quadrupole) approximation is accurate at a level that
allows a set of specific templates to be used. 

Relativistic corrections to the time frequency behavior are typically
$<$10\%\ of the Newtonian contribution and can be extracted from the
signal to high accuracy.  A great deal of phenomenological work has
been done to determine the number and range of templates required, the
efficiency for extraction of signals in background noise, {\it etc.}.
The results indicate that the range of anticipated neutron star
parameters can be covered with a manageable number of templates.  The
final coalescence of NS/NS systems will yield information sensitive to
the equation of state of nuclear matter, however this part of the
spectrum is typically at frequencies 1 kHz or higher where the shot
noise in the interferometers are a serious limitation.
If such sources are
observed, however, future configurations for the LIGO interferometer
promise to yield the ability to have improved sensitivity in a
narrower bandwidth which can be used for these studies.

The expected rate of such events is expected to be a few per year
within about 200 Mpc from neutron star pairs.  The rate is more
uncertain for black hole pairs, but due to the heavier masses they
make a large signal which will allow a deeper search into the Universe
for a given LIGO sensitivity.

{\bf Burst Signals}: The gravitational collapse of stars
(e.g. supernovae) will lead to emission of gravitational radiation.
Type I supernovae involve white dwarf stars and are not expected to
yield substantial emission.  However, Type II collapses can lead to
strong radiation, if the core collapse is sufficiently
non-axisymmetric. Estimates of the strengths indicate detection might
be possible out to the Virgo Cluster, which would yield rates of one
or more per year.  However, the gravitational wave signal depends on
the non axisymmetric component of the collapse and this is not well
determined by calculation.  Detection within or near or galaxy,
however, seems assured even if the collapse is highly symmetrical.
Calculations indicate that the signal from convectively unstable
neutron starts during the first second or so of its life should be
detectable in the frequency band of LIGO from sources throughout our
galaxy.

The detection of signals from supernovae is a challenge because the
waveforms are not well determined.  The duration and general
characteristics should allow identifying such burst signals with a
generic search for bursts, and assurance of detection will require
identifying burst like signals in coincidence from multiple
interferometers.  In addition, steps are underway to correlate signals
from the large neutrino detectors and similarly, the gravitational
wave signal can be coincidenced with these neutrino signals.

{\bf Periodic Signals:} Radiation from rotating non-axisymmetric neutron
stars will produce periodic signals in the detectors. The
gravitational wave frequency is twice the rotation frequency, which is
typically within the LIGO sensitivity band for known neutron
stars. Neutron stars spin down partially due to emission of
gravitation waves.  Searches for signals from identified neutron stars
will involve tracking the system for many cycles, taking into account
the Doppler shift for the motion of the Earth around the Sun, and
effects of spin-down of the pulsar. Both targeted searches for known
pulsars and general sky searches are anticipated.

{\bf Stochastic Signals:}
 Signals from gravitational waves emitted in the
first instants of the early universe ($t \sim 10^{-43}$ sec) can be detected
through correlation of the background signals from two or more
detectors. Some models of the early Universe can result in detectable
signals.  Observations of this early Universe gravitational radiation
would provide an exciting new cosmological probe.

\section{The LIGO Facilities}

The LIGO facilities at Hanford, WA and Livingston, LA each have a 4 km
``L'' shaped vacuum enclosure, which is 1 meter in diameter.  Vacuum is
required to reduce scattering off residual molecules that bounce off
the walls and are modulated by the small shaking of the vacuum walls
modulating this background.  In addition, we have installed baffles on
the walls to reduce scattering.  The large diameter of the tube is
both to minimize scattering and to provide the ability to house
multiple interferometers within the same facility.  Each facility has
a 4 km interferometer with test masses housed in vacuum chambers at
the vertex and the ends of the L shaped arms.  At Hanford, there will
also be a 2 km interferometer implemented in the same vacuum system,
allowing a triple coincidence requirement.  The overall vacuum system
is capable of achieving pressures of $10^{-9}$ torr.  We presently have
both arms at each site installed and they are vacuum tight.  We are
beginning to bake the tube to reach the desired high vacuum.  We
expect to have the entire vacuum system complete, all control systems
operational and at high vacuum before the end of 1999.

The initial detector for LIGO is a Michelson interferometer with a
couple of special features:

The arms are Fabry-Perot cavities to increase the sensitivity by
containing multiple bounces and effectively lengthening the
interferometer arms.  The number of bounces is set to not exceed half
the gravitational wave wavelength ($\sim$ 30 bounces).  The
interferometers are arranged such that the light from the two arms
destructively interferes in the direction of the photodetector, thus
producing a dark port.  However, the light constructively interferes
in the direction of the laser and this light is ``re-used'' by placing a
recycling mirror between the laser and beam splitter.  This mirror
forms an additional resonant cavity by reflecting this light back into
the interferometer, effectively increasing the laser power and thereby
the sensitivity of the detector.

Much work has been done over the past decade to demonstrate this
arrangement and the detailed techniques and required sensitivities in
smaller scale laboratory prototypes.  This includes experiments on a
40 m prototype interferometer at Caltech, which is a scale model of
LIGO, which has provided an excellent test bed to study sensitivity,
optics, controls and even some early work on data analysis, noise
characterization, {\it etc.}.  We also have built a special interferometer
(PNI) at MIT, which has successfully demonstrated our required phase
sensitivity, the limitation to the sensitivity at high frequencies.

\begin{figure}[ht]
\centerline{\epsfysize=14cm\epsfbox{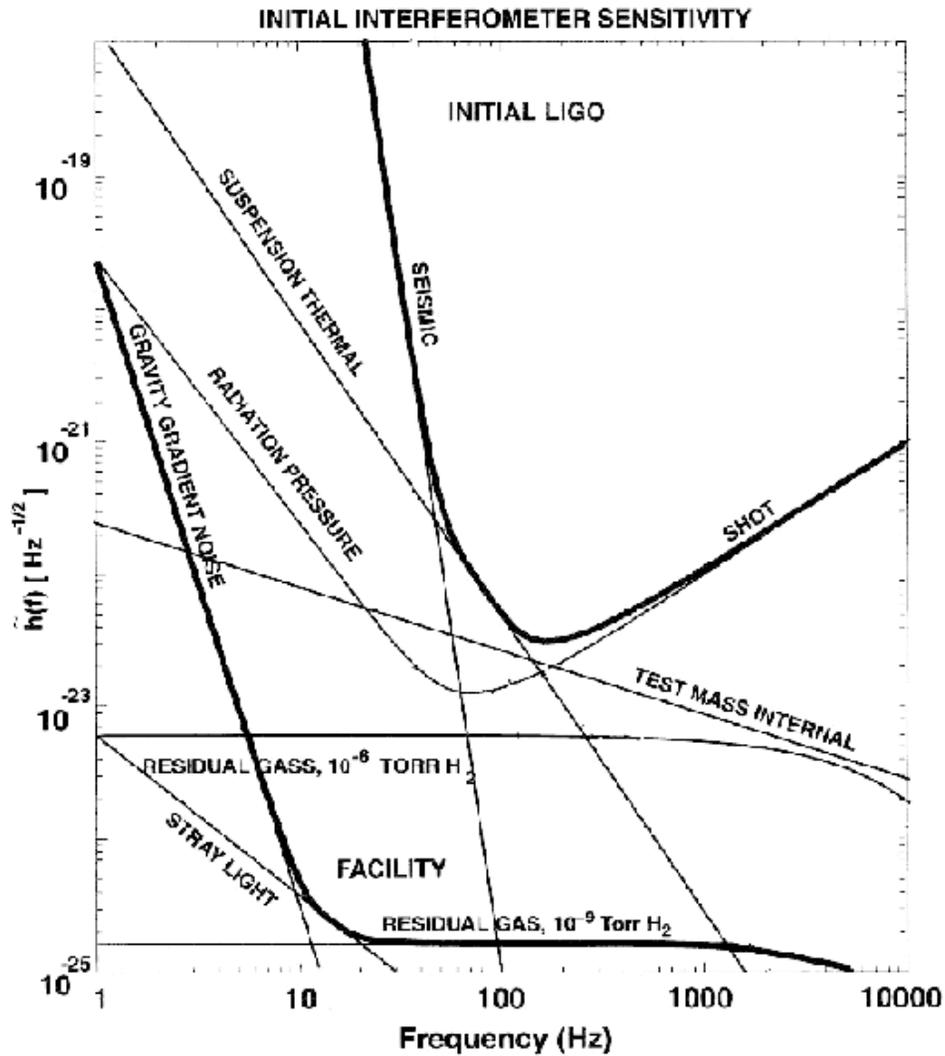}}
\smallskip
\caption[]{\label{fig5}\small
           The limiting noise curves are shown for the design of the
initial LIGO interferometers.  Note that at low frequency the limiting
source is seismic, at middle frequencies it is suspension thermal and
at the highest frequencies it is shot noise.  Some of the other
sources of noise that must be kept under control at a lower level are
shown demonstrating the room for improvement from the limiting noise
sources.}
\end{figure}

LIGO is limited in practice by three noise sources,
as illistrated in Figure \ref{fig5}:

At low frequencies ($\sim$ 10 Hz to 50 Hz), the limitation in sensitivity is
set by the level of seismic noise in the system.  We employ a seismic
isolation system to control this noise that consists of a four-layer,
passive vibration isolation stack having stainless steel plates
separated by constrained-layer damped springs.  This system is
contained within large vacuum chambers.  The stack supports an optical
platform from which the test mass is suspended.  The combination of
the seismic isolation stack and the test mass suspension give an
isolation from ground motion at the relevant frequencies of about 10
orders of magnitude.  The possibility of a more elaborate isolation
system and/or the addition of active isolation exists for the future,
as well as improved suspension systems.  Improvements in this area are
planned early in the future improvement program of LIGO.  

In the
middle range of frequencies ($\sim$ 50 Hz to 200 Hz) the principle effect
limiting the sensitivity is thermal noise.  This noise comes partially
from the suspension system, where there are violin wire resonances
from the steel suspension fibers.  However the principal noise source
is from the vibrational modes of the test masses.  This noise is
reduced by the choice of test masses, presently fused silica, to have
a very high Q thereby dissipating most of its noise out of our
frequency band.  The test masses will also improve in the future by
using higher Q fused silica, better bonding techniques for the wires,
and perhaps even new materials for the test masses, like sapphire.

At the higher frequencies (200 Hz to 5 kHz) the main limitation comes
shot noise and the sensitivity is limited by the power of the laser
(or the effective photostatistics).  The initial laser (Nd:YAG) is
designed and produced for LIGO, using a master oscillator/power
amplifier configuration, which yield a 10 watt high quality output
beam.  We also have developed a system to pre-stabilize this laser in
power and frequency.  Again, we expect to incorporate higher power
lasers in the future as they become available.

\begin{figure}[ht]
\centerline{\epsfysize=16cm\epsfbox{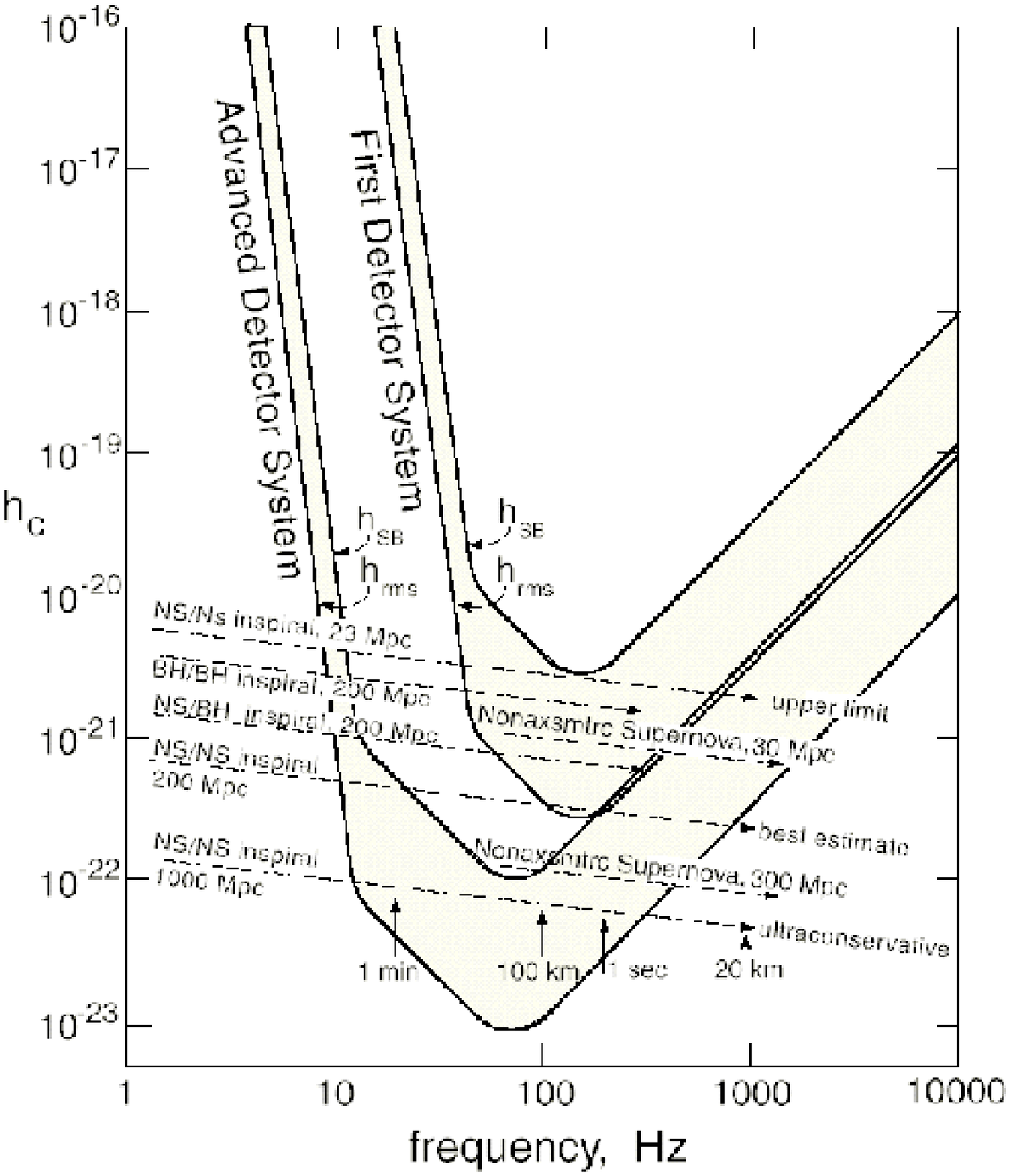}}
\smallskip
\caption[]{\label{fig6}\small
           Sensitivity of initial LIGO interferometers ($\sim$2002-2004)
and advanced LIGO interferometers ($\sim$2007-2010) with signal strengths
for various sources shown.  The lower line $h_{rms}$ 
is the noise expected
noise level and $h_{SB}$ represents approximate level for a signal/noise
allowing reliable signal detection.}
\end{figure}

The expected sensitivity of the inital LIGO detectors
and the advanced LIGO II detectors is shown in Figure \ref{fig6},
in comparison with signal strengths for various sources.

\section{Conclusions and Prospects}

The LIGO interferometer parameters have been chosen such that our
initial sensitivity will be consistent with estimates needed for
possible detection of known sources.  Although the rate for these
sources have large uncertainty, improvements in sensitivity linearly
improve the distance searched for detectable sources, which increases
the rate by the cube of this improvement in sensitivity. So,
anticipated future improvements will greatly enhance the physics reach
of LIGO and for that reason a vigorous program for implementing
improved sensitivities is integral to the design and plans for LIGO.

We are now entering into the final year of the construction of the
LIGO facilities and initial detectors.  We have formed the scientific
collaboration that will organize the scientific research on LIGO.
This collaboration already consists of more than 200 collaborators
from 22 institutions.  By early in the next millennium we will turn on
and begin the commissioning of these detectors.  We anticipate that we
will reach a sensitivity of $h \sim 10^{-21}$
by the year 2002.  At that
point, we plan to enter into the first physics data run ($\sim$ 2 years) to
search for sources.  This will be the first search for gravitational
waves with sensitivity where we might expect signals from known
sources.  Following this run in 2004, we will begin incremental
improvements to the detector interleaved with further data runs.  We
expect to reach a sensitivity of $h \sim 10^{-22}$
within the next 10 years,
making direct detection of gravitational waves within that time frame
reasonably likely.


\begin{references}  

\bibitem{ref1}
R.A.~Hulse and J.H.~Taylor. {\it Astrophys. J.}, 324 (1975); and
J.H.~Taylor, {Rev~Mod.~Phys.} 66 (1994).

\bibitem{ref2} A.~Abramovici {\it et al}, {\it Science}, 256, (1992).

\end{references}
\end{document}